\documentclass[iop,showkeys]{emulateapj}
\usepackage{apjfonts}
\shorttitle{MAGNETIC RESTRUCTURING IN TWO HOMOLOGOUS FLARES}
\shortauthors{LIU ET AL.}
\submitted{Received 2014 July 2; accepted 2014 September 22; published 2014 --}
\journalinfo{ApJ accepted 09/22/14}
\usepackage[breaklinks=true,setpagesize=false,bookmarks=false,colorlinks=true,linkcolor=blue,citecolor=blue,urlcolor=blue]{hyperref}
\newcommand{\sm}{$\sim$}
\newcommand{\goes}{\textit{GOES}}
\newcommand{\kms}{km~s$^{-1}$}
\newcommand{\hmi}{Helioseismic and Magnetic Imager}
\newcommand{\aia}{Atmospheric Imaging Assembly}
\newcommand{\Hsi}{\textit{Reuven Ramaty High Energy Solar Spectroscopic Imager}}
\newcommand{\hsi}{\textit{RHESSI}}
\newcommand{\fermi}{\textit{Fermi}}

\newcommand{\dg}{$^{\circ}$}
\newcommand{\Sdo}{\textit{Solar Dynamics Observatory}}
\newcommand{\sdo}{\textit{SDO}}

\begin{document}

\title{THREE-DIMENSIONAL MAGNETIC RESTRUCTURING IN TWO HOMOLOGOUS SOLAR FLARES IN THE SEISMICALLY ACTIVE NOAA AR 11283}

\author{Chang Liu\altaffilmark{1}, Na Deng\altaffilmark{1}, Jeongwoo Lee\altaffilmark{1,2}, Thomas Wiegelmann\altaffilmark{3}, Chaowei Jiang\altaffilmark{4,5}, Brian R. Dennis\altaffilmark{6}, Yang Su\altaffilmark{7},\\Alina Donea\altaffilmark{8}, and Haimin Wang\altaffilmark{1}}
\affil{$^1$~Space Weather Research Laboratory, New Jersey Institute of Technology, University Heights, Newark, NJ 07102-1982, USA; \href{mailto:chang.liu@njit.edu}{chang.liu@njit.edu}}
\affil{$^2$~Department of Astronomy and Space Science, Chungnam National University, Daejeon 305-764, Korea}
\affil{$^3$~Max-Planck-Institut f{\"u}r Sonnensystemforschung, Justus-von-Liebig Weg 3, 37077 G{\"o}ttingen, Germany}
\affil{$^4$~Center for Space Plasma and Aeronomic Research, The University of Alabama in Huntsville, Huntsville, AL 35805-1911, USA}
\affil{$^5$~SIGMA Weather Group, State Key Laboratory for Space Weather, Center for Space Science and Applied Research, CAS, Beijing 100190, China}
\affil{$^6$~Code 671, NASA Goddard Space Flight Center, Greenbelt, MD 20771-2400, USA}
\affil{$^7$~IGAM-Kanzelh\"{o}le Observatory, Institute of Physics, University of Graz, Universit\"{a}tsplatz 5, A-8010 Graz, Austria}
\affil{$^8$~Centre from Stellar and Planetary Astrophysics, School of Mathematical Sciences, Monash University, Melbourne, Victoria 3800, Australia}

\begin{abstract}
We carry out a comprehensive investigation comparing the three-dimensional magnetic field restructuring, flare energy release, and the helioseismic response, of two homologous flares, the 2011 September 6 X2.1 (FL1) and September 7 X1.8 (FL2) flares in NOAA AR 11283. In our analysis, (1) a twisted flux rope (FR) collapses onto the surface at a speed of 1.5~\kms\ after a partial eruption in FL1. The FR then gradually grows to reach a higher altitude and collapses again at 3~\kms\ after a fuller eruption in FL2. Also, FL2 shows a larger decrease of the flux-weighted centroid separation of opposite magnetic polarities and a greater change of the horizontal field on the surface. These imply a more violent coronal implosion with corresponding more intense surface signatures in FL2. (2) The FR is inclined northward, and together with the ambient fields, it undergoes a southward turning after both events. This agrees with the asymmetric decay of the penumbra observed in the peripheral regions. (3) The amounts of free magnetic energy and nonthermal electron energy released during FL1 are comparable to those of FL2 within the uncertainties of the measurements. (4) No sunquake was detected in FL1; in contrast, FL2 produced two seismic emission sources S1 and S2 both lying in the penumbral regions. Interestingly, S1 and S2 are connected by magnetic loops, and the stronger source S2 has weaker vertical magnetic field. We discuss these results in relation to the implosion process in the low corona and the sunquake generation.
\end{abstract}


\onlinematerial{color figures, animations}

\section{INTRODUCTION}\label{introduction}
There is often a coincidence among flare emissions, launch of the coronal mass ejections (CMEs), and the impulsive seismic waves (``sunquakes'') embarking on and beneath the Sun's surface, and they are usually considered to be related to each other. However, the transport of energy and momentum into the interior from the solar atmosphere is not yet fully understood. Since solar eruptions are generally believed to be a result of magnetic reconnection in the corona \citep[e.g.,][]{forbes13,su13}, changes in the magnetic field due to coronal field restructuring must be considered not only in studying flares/CMEs but also in understanding the coupling between the corona and the solar interior \citep{hudson11}. Efforts in this regard were originally initiated from the theoretical viewpoint. In particular, the concept of ``implosions'' in coronal transients (cf. outward explosions of CMEs) was put forward to predict an inward contraction of coronal field that would occur simultaneously with the magnetic energy release \citep{hudson00}. Clear evidence of implosions during the flare impulsive phase has recently been gathered from time sequences of high-resolution coronal images \citep[e.g.,][]{Liur+implosion09,liur12b,gosain12,simoes13}. The implosion picture may be considered in line with the reconnection model of two current-carrying flux loops, which produces two new current-carrying flux systems with one moving closer to the surface and another moving upward as part of CMEs \citep{melrose97}. This kind of loop-loop interaction is depicted in the tether-cutting reconnection model for sigmoidal active regions (ARs) \citep[e.g.,][]{moore01,liu07b,liur10,liu13}. The downward drop of coronal currents after eruptions has also been indicated by comparing the pre- and postflare coronal field models reconstructed using the nonlinear force-free field (NLFFF) extrapolation technique \citep{liu12,sun12}.

A direct consequence of coronal implosions would be a more horizontal (i.e., more inclined to the surface) configuration of the photospheric magnetic field. This is supported by the finding that the transverse magnetic field around the magnetic polarity inversion line (PIL) at the center of the flaring region often exhibits a rapid and persistent enhancement immediately after flares/CMEs, as revealed by vector magnetograms of either ground-based or space-borne instruments \citep[e.g.,][]{wang10,liu12,wangs12,wangs12b,sun12,petrie13}. Intuitively, such an inward collapse of the central magnetic field may be accompanied by an upward turning of the peripheral field in the outer areas of flaring regions. At these locations the photospheric transverse field would subsequently decrease, as sometimes reported \citep{li11,sun12}. Since sunspot penumbrae in white light (WL) indicate an inclined magnetic field structure, it is remarkable to note that the above surmised collapse-and-turning scenario of flaring magnetic field (as simply but attractively illustrated in \citealt{liu05}) is well corroborated by the observed flare-related strengthening (decay) of central (outer) sunspot penumbra of $\delta$ spots \citep{wang04a,liu05,deng05,chen07,deng11,wang12b,wang13}. 

Importantly, the change of photospheric magnetic field at the flare core region to a more horizontal state implies an upward Lorentz-force change exerting on the outer atmosphere, which is balanced by an equal and opposite Lorentz-force change acting on the photosphere and below \citep{hudson08,fisher12}. The latter is described as a ``back reaction'' caused by the coronal field reconfiguration \citep{hudson08}. It is asserted that the integration of the upward Lorentz-force change over its change period should produce an upward impulse driving the erupting CMEs; the conservation of momentum also signifies an equal, downward-moving impulse, which is applied into the solar interior and could supply enough energy for seismic waves. However, the flare-CME-sunquake relation as implied above is not yet fully understood. More particular conditions may have to be fulfilled to generate a CME or a sunquake since not every flare is associated with them.

Sunquakes are produced in a response of the low solar atmosphere to the localized hydrodynamic impact of flares/CMEs. As a result, acoustic waves are generated, and they travel into the interior and refract back to the solar surface (for reviews, see \citealt{donea11} and \citealt{sasha14b}). Although circular ripples emanating from a flare site were discovered more than a decade ago \citep{sasha98}, the physical origin of this wave phenomenon is still unknown. The main driving mechanisms of sunquakes proposed thus far include precipitation of high-energy electrons or protons \citep[e.g.,][]{sasha07,zharkova07}, back-warming radiation heating related to WL emission \citep[e.g.,][]{donea06a,donea11}, and changes of magnetic field \citep[e.g.,][]{hudson08,martinez09,zharkov11,fisher12,alvarado12}. As the production of seismic activities could involve multiple layers of the solar atmosphere, study of sunquakes will advance the current understanding of solar flare physics.

We note that the coronal implosion scenario includes multiple components for back reaction, and the role of of each component has been studied separately. In this paper, we carry out a comprehensive investigation of the homologous 2011 September 6 X2.1 and September 7 X1.8 flares, including the coronal field restructuring using the NLFFF modeling, photospheric magnetic field and sunspot structure changes, energy release during flares, and the helioseismic response to the flare impact. We consider it instructive to inspect the differences between these two events, because such a comparative study could be important to better explore and understand the role and relationship of different phenomena associated with flares. The plan of this paper is as follows: in Section~\ref{data}, we first provide an events overview and then introduce the observations and the data processing procedure. In Section~\ref{result}, we describe the observational and model results and point out their implications. Major findings are summarized and discussed in Section~\ref{summary}.

\section{EVENTS OVERVIEW AND DATA PROCESSING}\label{data}
\begin{figure}
\epsscale{1.1}
\plotone{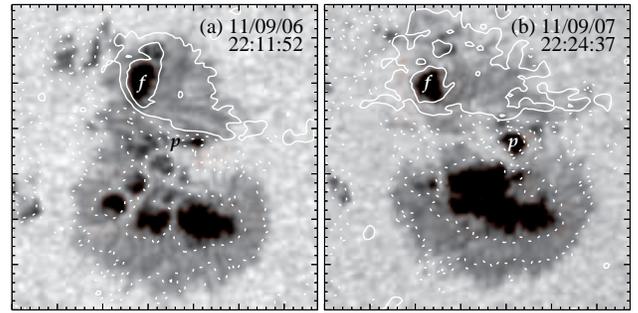}
\caption{HMI continuum intensity images right before the X2.1 flare on 2011 September 6 (a) and the X1.8 flare on September 7 (b), showing the evolution of NOAA AR 11283 in one day. The solid and dotted contours represent positive and negative magnetic field, respectively, at levels of $\pm$200 and $\pm1000$~G. p and f indicate the preceding and following spots, respectively. The images have a field of view of 67\arcsec~$\times$~67\arcsec\ (same as Figures~\ref{f1}, \ref{f3}(a)--(d), and \ref{f7}(a)--(f)). \label{f0}}
\end{figure}

The source NOAA active region (AR) 11283 emerged on the east limb on 2011 August 30. No major flares occurred until September 6, when the region appears in the $\beta\gamma$/$\beta\delta$ configuration containing one $\delta$ spot (see Figure~\ref{f0}). The September 6 X2.1 and September 7 X1.8 flares are the only X-class flares that occurred in this AR, and they make up a pair of homologous flares/CMEs due to similar locations and emission signatures in multiple wavelengths. Several studies of these flares have been conducted. \citet{wangs12b} and \citet{petrie13} studied the evolution of the photospheric vector magnetic field related to the flares. They found a significant increase of the horizontal field component that corresponds to a notable Lorentz-force perturbation. These changes are considered rapid (in \sm30 minutes) compared to the long-term evolution of the AR field on a timescale of days. \citet{jiang14} examined the buildup process and instability condition of the AR sigmoid before the X2.1 flare using a NLFFF extrapolation method. A realistic initiation of the eruption was then achieved by \citet{jiang13} based on an MHD model. \citet{feng13} analyzed the energy budget of the first event, and concluded that this X2.1 flare and the associated CME had similar total energies. The sunspot rotation before the onset of the X2.1 flare was also observed and linked to the event onset \citep{ruan14}. In particular, the second flare (X1.8) produces a pronounced sunquake as reported by \citet{zharkov13}; however, no clear seismic signatures were detected during the earlier X2.1 flare, and these homologous flaring events seem to have similar emission patterns in WL and X-rays \citep{xu14}.

\begin{figure*}
\epsscale{1.17}
\plotone{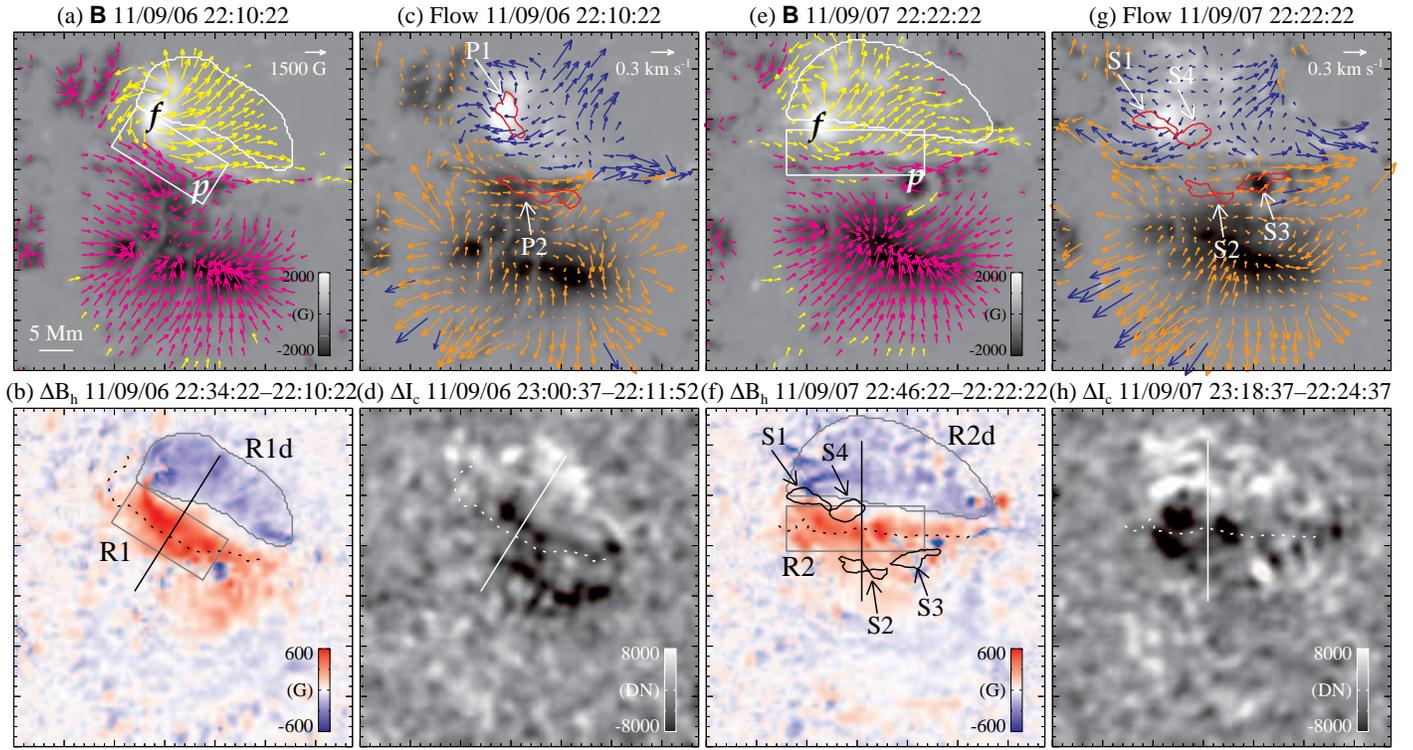}
\caption{Preflare HMI vertical magnetic field superimposed with arrows illustrating horizontal field vectors (a and e) and plasma flows (c and g). For clarity, arrows in negative and positive fields are coded in different colors. The flare-related changes of the horizontal field (b and f) and intensity (d and h) are presented. The intensity difference images are smoothed by a window of 2.5\arcsec~$\times$~2.5\arcsec. The running difference images of intensity near the HXR peak are contoured at 3800 DN in (c) for the 2011 September 6 X2.1 flare (22:18:37~UT frame minus 22:17:52~UT frame), and in (f) and (g) for the September 7 X1.8 flare (22:36:37~UT frame minus 22:35:52~UT frame), showing flare impacts P1--P2 and S1--S4, respectively; see Section~\ref{sunquake} for details. In the lower panels, the solid straight lines indicate the bottom side of the rotating vertical slice used in Figure~\ref{f3}. The dotted lines show the main segment of the PIL.\vskip 5mm \label{f1}}
\end{figure*}

In this analysis, full-disk photospheric vector magnetograms are supplied by the \hmi\ (HMI; \citealt{schou12}) on board the the \Sdo\ \cite[\sdo;][]{pesnell12}. The vector field data are derived using the Very Fast Inversion of the Stokes Vector algorithm \citep{borrero11}, and the minimum energy method \citep{metcalf94,leka09a,metcalf06} is used to resolve the 180$^{\circ}$ azimuthal ambiguity. We acquired vector magnetograms of AR 11283 that were prepared by the HMI team for special campaigns for selected ARs\footnote{\url{http://jsoc.stanford.edu/jsocwiki/ReleaseNotes2}}. The images are remapped using Lambert (cylindrical equal area) projection centered on the midpoint of the AR, which is tracked at the Carrington rotation rate. The observed fields are also transformed to heliographic coordinates \citep{gary_hagyard90} to remove the projection effect. The vector data analyzed have a pixel scale of \sm0.5\arcsec\ and a cadence of 12 minutes, spanning from 20:22~UT on 2011 September 6 to 00:46~UT on September 8 throughout both the X2.1 and X1.8 flares. The former (latter) started at 22:12~UT (22:32~UT), peaked at 22:20~UT (22:38~UT), and ended at 22:24~UT (22:44~UT) on September 6 (7) in \goes\ 1--8~\AA\ flux.

Modeling of the coronal field was carried out using NLFFF extrapolations. The net force and torque in the observed photospheric field were first minimized by a preprocessing procedure in order to obtain the chromosphere-like data that meet the force-free condition \citep{wiegelmann06}. The ``weighted optimization'' method \citep{wheatland00,wiegelmann04} was then applied to the preprocessed photospheric boundary to extrapolate the magnetic field towards coronal heights. This approach minimizes an integrated joint measure $L$ over the volume $V$, which comprises the normalized Lorentz force, the field divergence, and treatment of the measurement errors (especially for the transverse field $B_{\rm T}$) \citep{wiegelmann10}:

\begin{eqnarray}
L = \int_{V} w_f \frac{|( \nabla \times {\bf B}) \times {\bf B}|^2}{B^2} \ + \ w_d \,|\nabla \cdot {\bf B}|^2 \, {\rm d}^3V
\nonumber\\
+ \ \nu \int_{S} ({\bf B} - {\bf B}_{\rm obs})\cdot{\bf W}\cdot({\bf B} - {\bf B}_{\rm obs}) \ {\rm d}^2S \ ,
\end{eqnarray}

\noindent where ${\bf B}$ is the magnetic field vector, $w_f$ and $w_d$ are weighting functions, $\nu$ is a Lagrangian multiplier that controls deviations between the model and observed (${\bf B}_{\rm obs}$) fields, and ${\bf W}$ is a space-dependent diagonal matrix incorporating errors. For optimal results, we set $\nu=0.001$, the vertical ($z$) component of ${\bf W}$ to 1, and the horizontal components ($x,y$) of ${\bf W}$ to $B_{\rm T}/{\rm max}(B_{\rm T})$ \citep{wiegelmann12}. Using 2~$\times$~2 rebinned magnetograms covering the entire AR, the field extrapolation was performed within a box of 300~$\times$~256~$\times$~256 uniform grid points, which corresponds to about 219~$\times$~187~$\times$~187~Mm$^3$. Note that the bottom boundary is assigned a uniform altitude of 0.73~Mm (1 pixel) in this study as a result of the preprocessed bottom boundary. For the 2 hrs period before and after both the X2.1 and X1.8 flares, the calculation was conducted at the full cadence (12 minutes); between the flares (from 00:22 to 20:22~UT on 2011 September 7), a 1 hr cadence was adopted. To quantitatively evaluate the model quality, we resorted to a current-weighted (CW) average of sine metric $\langle{\rm CW}$sin$\theta\rangle$ where $\theta$ is the angle between ${\bf B}$ and the electric current ${\bf J}$, and a mean absolute fractional flux change metric $\langle|f_i|\rangle=\langle|(\nabla \cdot{\bf B})|/(6|{\bf B}|/\Delta x)\rangle$ where $\Delta x$ is the grid spacing \citep[e.g.,][]{wheatland00,schrijver08}. These metrics have a zero value given a perfectly force-free and divergence-free field. For the inner computation box (excluding the buffer boundary of 32 pixels in the lateral and top) of our 63 NLFFF models, $\langle{\rm CW}$sin$\theta\rangle = 0.26 \pm 0.02$ and $\langle|f_i|\rangle \times 10^4 = 5.4 \pm 0.9$, which are typical for this algorithm and suggest moderately satisfactory solutions.

Evolution of the AR field can be revealed by surface flows, which were traced with the differential affine velocity estimator for vector magnetograms \citep[DAVE4VM;][]{schuck08}. A window size of 19 pixels for feature tracking was set according to former studies \citep[e.g.,][]{liuy13}. HMI also produces 45~s cadence continuum intensity images, Dopplergrams, and line-of-sight (LOS) magnetograms. The WL emission of flares as well as sunspot structure variation were examined using intensity images. A high-resolution (0.1\arcsec) TiO (a proxy for continuum at 7057~\AA) image taken by the New Solar Telescope \citep[NST;][]{goode10,cao10} at Big Bear Solar Observatory (BBSO) further helps to recognize the sunspot WL structure in detail. Based on the Dopplergrams, the acoustic sources of sunquakes were probed by applying the helioseismic holography technique \citep[e.g.,][]{donea99,lindsey00,donea05}, which reconstructs the acoustic emission by producing phase-coherent snapshots of acoustic egression power. Time profiles of the integration of the source area on the snapshots then give the temporal evolution of sunquake sources.

The hard X-ray (HXR) emission of the 2011 September 6 X2.1 flare was registered by the \Hsi\ \citep[\hsi;][]{lin02}, and the HXR images and spectra were investigated by \citet{feng13}. The 2011 September 7 X1.8 flare occurred during \hsi\ night but it was observed by the \fermi\ Gamma-Ray Burst Monitor \citep[GBM;][]{meegan09}, with data analysis made possible by the \fermi\ Solar Flare Observations facility\footnote{\url{http://hesperia.gsfc.nasa.gov/fermi_solar}}. \fermi/GBM produces HXR time profiles and spectra with no imaging capability. In addition, to identify coronal magnetic structures (e.g., filaments and hot coronal loops) related to eruptions, we used 304~\AA\ (He~{\sc ii}; 0.05~MK) and 94~\AA\ (Fe~{\sc xviii}; 6.3~MK) images taken by the \aia~(AIA; \citealt{lemen12}) on board \sdo.

\section{RESULTS AND ANALYSIS} \label{result}
In this section, we describe observational and model results obtained using all data sets, and discuss their significance. Specifically, we first present the flare-related changes of magnetic field on the surface, then proceed to the magnetic field restructuring in three dimension (3D) (more dynamic detail can be seen in an accompanying animation in the online journal\footnote{\url{http://web.njit.edu/~cl45/ms_2011090607/}}), and subsequently estimate the amount of the free magnetic energy released during the flares as well as that of nonthermal energy. Finally, we study seismic sources and their relation to flare emissions and magnetic field structure.

\begin{figure}
\epsscale{1.177}
\plotone{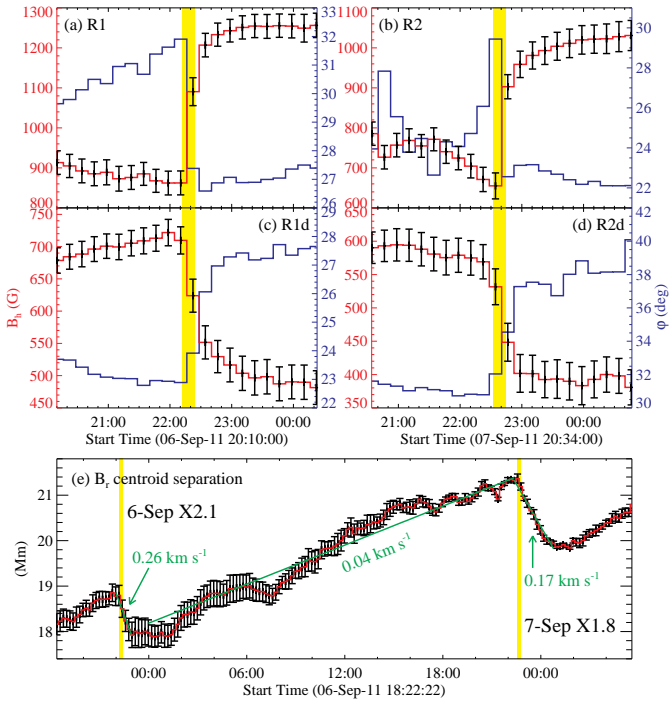}
\caption{Temporal evolution of the mean horizontal field $B_{\rm h}$ and inclination angle $\varphi$ in the regions R1 (a), R2 (b), R1d (c), and R2d (d), as denoted in Figure~\ref{f1}, and that of the flux-weighted centroid separation of the two magnetic polarities of the vertical field (e). The green lines are the linear fits to the data. The yellow shades indicate the duration of the X2.1 and X1.8 flares in \goes\ 1--8~\AA\ flux.\vskip 5mm \label{f2}} 
\end{figure}

\subsection{Surface Magnetic Field Changes} \label{surface}
Both the 2011 September 6 X2.1 and September 7 X1.8 flares are initiated around the highly sheared PIL of the AR (Figures~\ref{f1}(a) and (e)). In this area magnetic field concentrations with opposite polarity show a converging motion in the N-S direction and a diverging motion in the E-W direction, which are clearly tracked by DAVE4VM (Figures~\ref{f1}(c) and (g)). The main proceeding (p) and following (f) spots of the $\delta$ configuration gradually separate from each other from September 6 to 7, which eventually ``straightens'' the PIL (also see Figure~\ref{f0}). To better delineate the entire evolution and study the position change of the major photospheric magnetic field concentrations, we calculate the centroid (flux-weighted average) positions ($x_c, y_c$) of both polarities as $x_c = \sum x_iB_{{\rm r}(i)} / \sum B_{{\rm r}(i)}$ and $y_c = \sum y_iB_{{\rm r}(i)} / \sum B_{{\rm r}(i)}$, where ($x_i, y_i$) is the coordinate of pixels with an absolute vertical magnetic flux density $|B_{{\rm r}(i)}| \geq B_0$ \citep{wang06}. For this AR, $B_0$ is set to vary from 800~G to 1000~G for uncertainty estimation. The temporal evolution of the centroid separation between the main positive and negative fields is plotted in Figure~\ref{f2}(e), which demonstrates their separation motion at a speed of about 0.04~\kms\ at most times. It can also be obviously seen that the centroid separation decreases immediately after both flares (the vertical yellow shades) for a short time period, after which the long-term evolution trend is restored. Specifically, the separation of the two polarities is shortened by \sm0.85~Mm in 1 hr (at \sm0.26~\kms) following the earlier X2.1 flare, and by \sm1.4~Mm in 2.4 hrs (\sm0.17~\kms) following the later X1.8 flare. These temporary but rapid approaching of the two components of the $\delta$ spot are opposite to the overall shearing motion, implying the relaxation of the sheared magnetic field after the flares \citep{wang06} in accord with magnetic implosions \citep{hudson00}. Here the decrease of the centroid separation distance associated with the X1.8 flare is more significant (\sm1.6 times) compared to that associated with the X2.1 flare, implying a larger energy release in the former event.

Rapid changes of the horizontal field $B_{\rm h}$ are also found in both flares. A stepwise increase of $B_{\rm h}$ (colored red in the difference $B_{\rm h}$ maps in Figures~\ref{f1}(b) and (f)) is primarily observed within the elongated region lying along the central flaring PIL, as outlined using the box R1 (R2) (also see \citealt{petrie13}). There the mean $B_{\rm h}$ abruptly increases by \sm26\% in the X2.1 and by \sm38\% in the X1.8 flare (Figures~\ref{f2}(a) and (b)). Corresponding to the surface field change, the radial component of the Lorentz-force change acting on the surface and interior can be quantified as $\delta F_r = \frac{1}{8\pi} \int dA (\delta B_{\rm r}^2-\delta B_{\rm h}^2)$ \citep{fisher12}. Integrating over the region R1 (R2) yields a similar \textit{downward} $\delta F_r$ peaking at $2.5 \times 10^{22}$ ($2.1 \times 10 ^{22}$) dyne during the X2.1 (X1.8) flare. The increase of $B_{\rm h}$ around flaring PILs is a common feature of major flares \citep[e.g.,][]{wang10}, and could be closely related to implosions of the coronal field \citep[e.g.,][]{liu12}.

More interestingly, the central areas of $B_{\rm h}$ enhancement are surrounded by the ring-like regions with decreasing $B_{\rm h}$ (colored blue in Figures~\ref{f1}(b) and (f)). It is remarkable that the rings are not symmetric, but much more intense on the northern side in the fan-shaped sections R1d and R2d. The mean $B_{\rm h}$ in R1d (R2d) tends to decrease for about 0.5 hr, with a sharpest drop by 12\% (16\%) in the X2.1 (X1.8) flare (Figures~\ref{f2}(c) and (d)).

We note that (1) previous works are mainly focused on the increase of $B_{\rm h}$ in the flare core region, while the weakening of $B_{\rm h}$ in the peripheral areas has not been studied in detail. (2) The distinct asymmetry of the decreasing $B_{\rm h}$ region is unlikely an artifact due to the projection effect. This is because the flaring AR is considered not far from the disk center (with an orientation cosine factor of 0.4 at the X2.1 flare and 0.54 at the X1.8 flare), and that both events show a similar change pattern. Instead the asymmetric field change on the photosphere may connote an asymmetry in the 3D magnetic field structure and its evolution. This will be further discussed in the next section. (3) The ring-like change pattern of $B_{\rm h}$ is very reminiscent of similar features in the flare-related darkening (decay) of the inner (outer) penumbra \citep{liu05}, which are also observed in the present events. In the difference images before and after the flares (Figures~\ref{f1}(d) and (h); cf. Figure~\ref{f0}), we see the central dark region (mainly penumbral strengthening) encompassed by asymmetric white areas (penumbral decay), which well corresponds to the regions of $B_{\rm h}$ increase and decrease, respectively. The magnetic inclination angle relative to the horizontal plane $\varphi={\rm tan}^{-1}(|B_r|/B_h)$ also changes accordingly after both flares. It rapidly decreases by 14\% (23\%) in the R1 (R2) region, while increases gradually in the R1d/R2d regions by \sm18\% in 0.5 hr (Figures~\ref{f2}(a)--(d)). 

\begin{figure*}
\epsscale{1.177}
\plotone{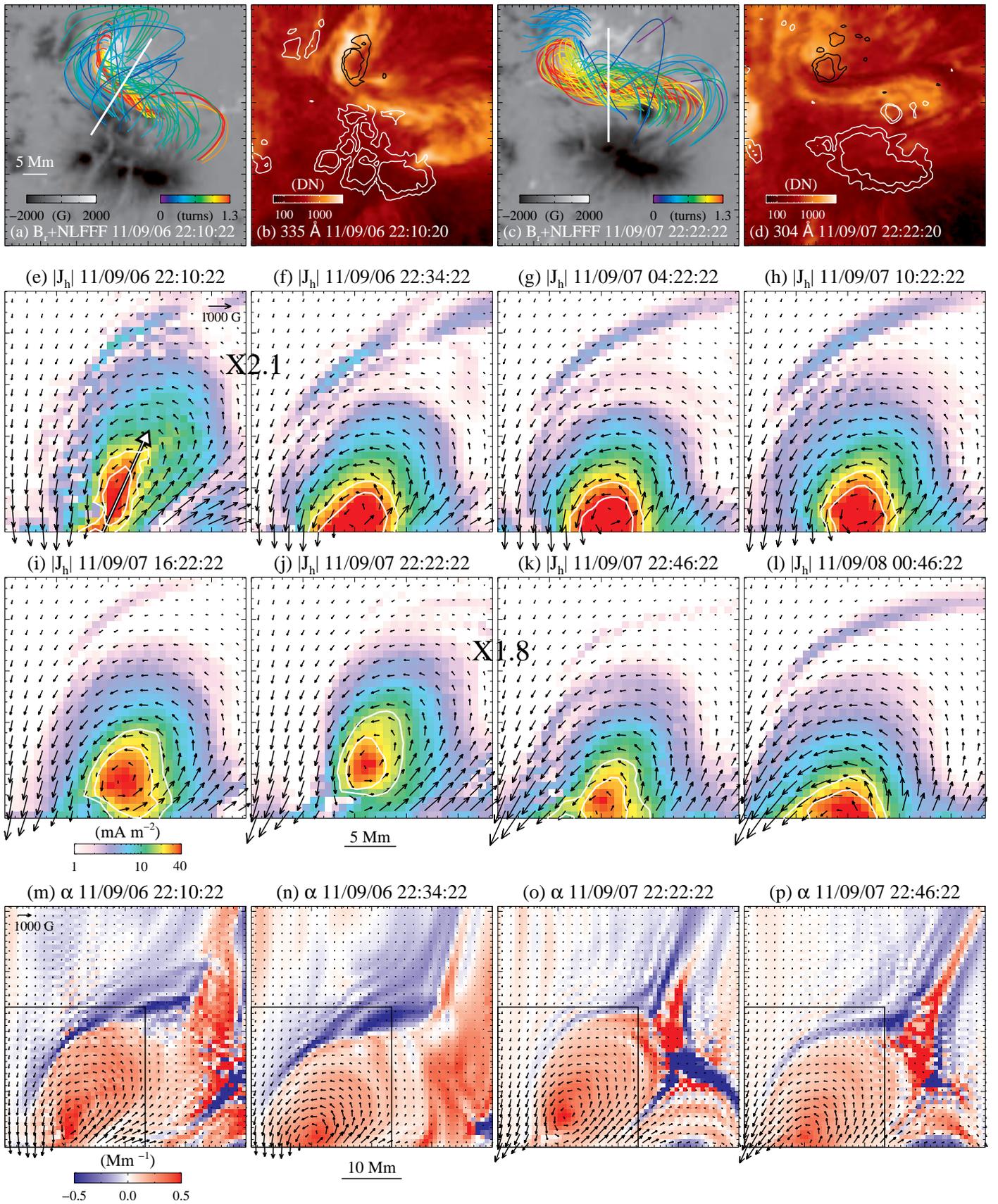}
\caption{\footnotesize{Preflare HMI vertical field images superimposed with selected NLFFF lines (a and c), in comparison with cotemporal AIA 304~\AA\ images (b and d) overplotted with vertical field contours at levels of $\pm$1000 and $\pm$1500~G. The thick white slit in (a) and (c) indicates the bottom of a rotating vertical slice of 23~$\times$~23~Mm$^2$, over which the distributions of $J_{\rm h}$ are plotted in (e)--(l) in logarithmic scale at selected instances. The white contours are at levels of 40\% and 60\% of the maximum current. The left bottom corner of the vertical slice corresponds to the southern end of the slit. The thick arrow in (e) indicates the FR orientation in the cross section. (m)--(p) shows the distribution of the force-free parameter $\alpha$ in a larger slice (40~$\times$~40~Mm$^2$) further extending northward, with the field of view of (e)--(l) denoted by the black box. The black arrows in (e)--(p) are the transverse field vectors in the vertical slices. See Section~\ref{vertical} for details.}  \label{f3}}
\end{figure*}

As will be shown below, we believe that the observed flare-related photospheric changes are an intrinsic manifestation of a reconfiguration of coronal fields. This view is consistent with the simulation of \citet{rempel11}, in which sunspot penumbrae form or become destructed within \sm0.5 hr after the boundary condition at the top is altered.

\subsection{3D Magnetic Field Restructuring} \label{vertical}
We first inspect the general coronal field structure right before the X2.1 and X1.8 flares using our NLFFF extrapolations. Selected field lines within the main $\delta$ spot region are plotted in Figures~\ref{f3}(a) and (c). These lines are colored according to the twist index $T_n=\frac{1}{4\pi}\alpha L = \frac{\mu_0}{4\pi} J_r L / B_r$ at their footpoints, where $J_r$ and $B_r$ are the vertical current and field on the photosphere, respectively, and $L$ is the field line length \citep{inoue11}. It is clearly shown that sheared and twisted field lines lie along the PIL and are embedded in less twisted envelope field. The former core fields possess an overall sigmoid structure tracing out a flux rope (FR), which is manifested by a sigmoidal filament observed in AIA 304~\AA\ (Figures~\ref{f3}(b) and (d)). Compared with the X2.1 flare on September 6, more highly twisted fields up to 1.9 turns are present right before the X1.8 flare on September 7, together with an apparently thicker filament. This may imply that the X1.8 flare could be associated with a more well-formed FR (see further discussions below). Detailed topological analysis of the FR as well as its instability condition are, however, out of the scope of this study (but see \citealt{jiang14}).

In the following, we investigate the time sequence of static NLFFF models assuming that the coronal field evolution can be described by equilibria in successive steps \citep[e.g.,][]{regnier06,wu09,liu12,sun12,jiang14,jing14}. This approach allows us to gain an insight into the relaxation of the nonpotentiality as well as the build-up process of magnetic energy between the eruptions. For our purpose of the comparative study of the 3D field restructuring associated with the X2.1 and X1.8 flares, we employ a vertical slice of 23~$\times$~23~Mm$^2$ cutting through the middle of the FR and concentrate on examining the evolution of magnetic field and electric current in this cross section. The bottom side of this slice on the surface is overplotted on Figures~\ref{f3}(a) and (c) and also Figures~\ref{f1}(e)--(h) as a thick straight line. The slice rotates counterclockwise at a speed of \sm1.33\dg~hr$^{-1}$ so that it maintains an orientation perpendicular to the central PIL. In Figures~\ref{f3}(e)--(l) (and the accompanying animation), we depict the distribution of the horizontal current $|{\bf J}_{\rm h}|$ superimposed with the transverse magnetic field within this rotating vertical slice at selected instances, including those immediately before and after the flares. Some snapshots of the force-free parameter $\alpha = \mu_0 {\bf J} \cdot {\bf B} / |{\bf B}|^2$ in a larger slice are also presented in Figures~\ref{f3}(m)--(p). We describe these results in detail as follows.

First, a spiral of field lines is evidently seen in the vertical cross section at most times, manifesting the poloidal flux of the FR. The system obviously carries free-magnetic energy as there exist significant field-aligned currents, with the maximum of $J_h$ found to be generally cospatial with the axis of the FR (i.e., the center of the spiral) \citep[e.g.,][]{canou10}. The FR is also twisted and the maximum $\alpha$ near its axis reaches about 0.57 (0.48) Mm$^{-1}$  right before the X2.1 (X1.8) flare. Also noticeably, the FR is not symmetric in the vertical cross section especially before both flares, as it leans northward at about 66\dg\ relative to the surface. This is illustrated by the thick long arrow in Figure~\ref{f3}(e), which starts from the PIL and passes through the $J_h$ maximum. The inclination of the FR towards north is mostly because that the magnetic flux (thus magnetic pressure) of the southern negative polarity is stronger than the northern positive polarity \citep{jiang14}. The overlying and neighboring field structure of the FR is delineated by the maps of the distribution of $\alpha$ in a wider field of view.

\begin{figure}
\epsscale{1.17}
\plotone{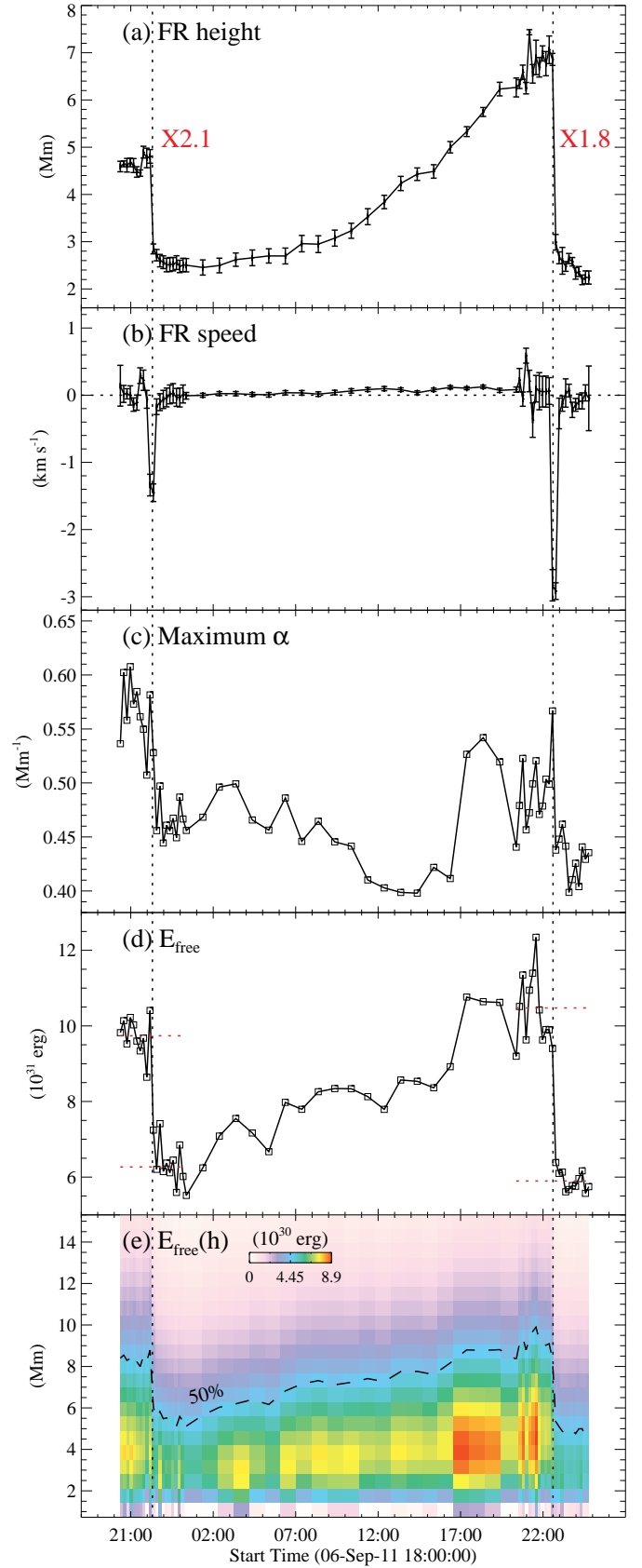}
\caption{Temporal evolution of the FR height (a), FR speed (b), maximum $\alpha$ near the FR axis (c), total free magnetic energy $E_{\rm free}$ (d), and $E_{\rm free}$ as a function of height (e). The vertical dotted lines indicate the peak times of the HXR emission of the X2.1 and X1.8 flares.\vskip 2mm \label{f4}}
\end{figure}

Second, similar to the ascending motion of the FR before the X2.1 flare \citep{jiang14}, the FR gradually rises between the occurrence times of the X2.1 flare on September 6 and the X1.8 flare on September 7 (Figures~\ref{f3}(f)--(j)). To aid in quantifying the FR evolution, we take the centroid of $J_h$ as a proxy for the FR axis and follow its temporal evolution projected onto the FR inclination direction. The statistical position error was estimated by varying the low threshold from 40\% to 60\% of each maximum $J_h$ (white contours in Figures~\ref{f3}(e)--(l)), as these contours cover a region similar to the observed filament width. The result in Figure~\ref{f4}(a) shows that the altitude of the FR axis gradually increases from 2.7~Mm at 22:34~UT on September 6 to 7.1~Mm at 22:22~UT on September 7. During 15--20~UT, the derived speed of this ascending motion has an average value of 0.09~\kms\ (see Figure~\ref{f4}(b)), which is comparable to the slow rising speed of filaments often measured before flares/CMEs \citep[e.g.,][]{liur12}. We speculate that this rising motion signifies the buildup process of the FR driven by the tracked converging flows in this AR (Figure~\ref{f1}(c)) \citep{van89}. Nevertheless, the center part of the filament is mainly observed to become thickened (cf. Figures~\ref{f3}(b) and (d)) without obvious motion. It is reasonable that in the present case the filament could be supported by the magnetic dips below the FR axis \citep{gilbert01}, which seem to only rise up to \sm3~Mm (Figure~\ref{f3}(j)). This displacement is hard to be detected considering the projection effect and image resolution. 

\begin{figure}
\epsscale{1.17}
\plotone{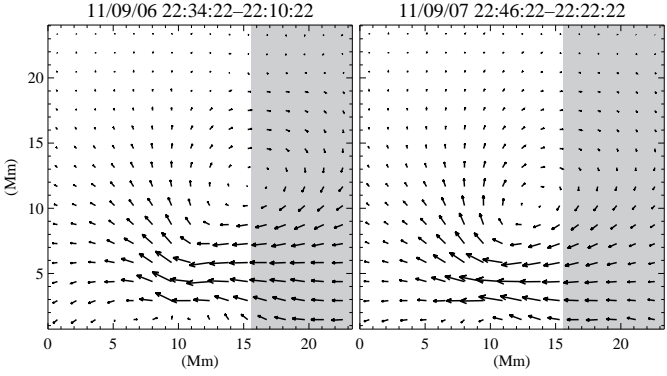}
\caption{Changes of the transverse field in the vertical slice (Figures~\ref{f3}(e)--(l)) from right before to right after the 2011 September 6 X2.1 (left) and September 7 X1.8 (right) flares, both showing an obvious leftward (southward) turning of magnetic field predominantly below about 10~Mm. The portion of the slice in gray corresponds to the northern end part of the slit drawn in the lower panels of Figure~\ref{f1} , which extends into the outer penumbral region.\vskip 5mm \label{f5}}
\end{figure}

Third, the FR appears to be attached to the photosphere before the X2.1 flare; in contrast, it has grown into a semi-circular structure with its main body apparently elevated off the surface (cf. Figures~\ref{f3}(e) and (j)). Most interestingly, with the release of the nonpotentiality as reflected by a sudden decrease of the force-free parameter $\alpha$ (Figure~\ref{f4}(c)), the FR undergoes an apparent downward collapse (inward contraction) toward the surface after both flares. This sharp evolution represents a major restructuring of the coronal field with several noticeable features. (1) The spiral of the helical core field lines survives the X2.1 flare but vanishes in 1 hr after the X1.8 flare (cf. Figures~\ref{f3}(g) and (l)). This alludes to the possibility that the former event is associated with a partial expulsion of the FR \citep{gibson06a,jiang14}, while the latter event is associated with a fuller FR eruption. (2) The collapse speed of the FR is about 1.5~\kms\ in the X2.1 flare and doubles to 3~\kms\ in the X1.8 flare (Figure~\ref{f4}(b)). Thus a more abrupt implosion process is very likely during the latter event, which is consistent with a larger percentage increase (decrease) of the horizontal field (inclination angle) on the photosphere. This is the first time that the speed of implosions of the core field is derived based on NLFFF extrapolation results. As a comparison, \citet{jing14} used NLFFF models to find an inward contraction of arcade fields overlying a flaring region, which had a similar speed (\sm1.8~\kms) and was also attributed to coronal implosions. Here we have to recognize that data points during the flares should be interpreted with caution, as flare emissions have potential impacts on the field measurement. (3) The FR drops in the X1.8 and X2.1 flares for a distance of about 4.2 and 2.1~Mm, respectively. The ratio of 2 in the vertical direction is consistent with that (a ratio of 1.6) of the decrease of the flux-weighted centroid separation between opposite polarities on the photosphere (Section~\ref{surface}). Due to the vastly different Alfv{\' e}n speeds, however, the time for the coronal field to relax to equilibrium takes a few 10 minutes, while that for the photospheric field takes 1--2 hrs. (4) Our attention is also drawn to the change of ambient fields adjacent to the FR. In Figure~\ref{f5}, we plot the differences of transverse field in the vertical slice between the immediate pre- and postflare states. It shows that almost all the fields below about 10~Mm have a clear leftward (southward) turning (also see the animation). This can be understood since the preflare FR has an inclined orientation towards the north due to the pressure imbalance, which might be alleviated after the flare energy release. The portion of the slice in gray in Figure~\ref{f5} corresponds to the outer penumbral field region, and a more vertical field there following the turning nicely agrees with the observed primary decrease of horizontal field and decay of penumbra in the R1d and R2d areas (Figure~\ref{f1}, lower panels). We conclude that all the above self-consistent results make a first comprehensive portray of the 3D coronal implosion picture in the low solar atmosphere.

\begin{figure}
\epsscale{1.17}
\plotone{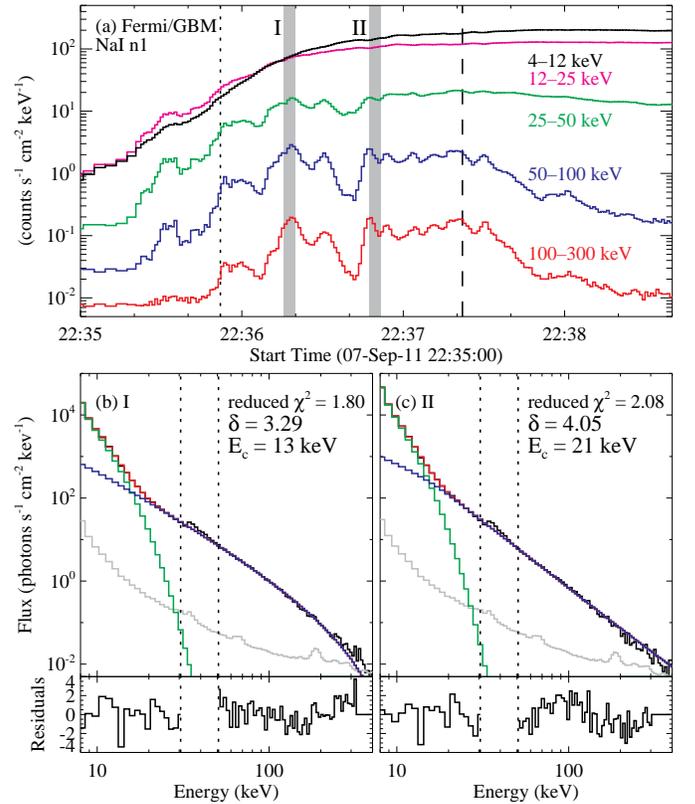}
\caption{(a) \fermi\ HXR light curves of the 2011 September 7 X1.8 flare. The dotted (dashed) line corresponds to 22:35:52~UT (22:37:22~UT). Spectral fits in the two 4~s intervals I and II (gray bands) around the HXR peak are shown in (b) and (c), using a thermal (green) and a nonthermal (blue) functions. Denoted fitting parameters include $\delta$, the electron power-law index and $E_c$, the low energy cutoff of the nonthermal electron spectrum. The 30--50~keV (between the dotted lines) range is excluded in the fitting; see Section~\ref{energy} for details.\vskip 5mm \label{f6}}
\end{figure}

\subsection{Flare Energy Release} \label{energy}
It is also worthwhile to calculate the energy content of the homologous X2.1 and X1.8 flares. Solar eruptions are believed to be powered by magnetic reconnection, which releases energy stored in the magnetic field. The upper limit of the available energy for energizing flares/CMEs is the free magnetic energy $E_{\rm free}$, which can be computed from

\begin{equation}
E_{\rm free}=\int_{V} \frac{B_{\rm N}^2}{8\pi} dV - \int_{V} \frac{B_{\rm p}^2}{8\pi} dV \ ,
\end{equation}

\noindent where $B_{\rm N}$ and $B_{\rm p}$ are the modeled NLFFF and potential field, respectively, based on the same observed boundary field. For all the time instances at which we construct NLFFF, potential field is also derived with a Fourier transform method \citep{alissandrakis81} using the vertical component of the HMI data. In Figures~\ref{f4}(d) and (e), we present the temporal evolution of the volume integrated $E_{\rm free}$ and the height profile of $E_{\rm free}$. Taking the difference of values averaged over 2 hrs before and after the flares (denoted by the horizontal red dotted lines), the released free magnetic energy $\Delta E_{\rm free}$ in the X2.1 (X1.8) flare is estimated to be $3.5~\times 10^{31}$ (4.6~$\times 10^{31}$) erg. The swift drop of $E_{\rm free}$ is well cotemporal with the peak times (the vertical dotted lines) of the flare HXR emissions (see the next section). Before the X1.8 flare, $E_{\rm free}$ accumulates mainly between 3 and 7~Mm, and decreases at nearly all heights afterward. A similar behavior is found before and after the X2.1 flare (also see \citealt{jiang14}).

Another form of energy that we are interested in is the energy imparted to nonthermal electrons $E_{\rm nonth}$. \citet{feng13} derived a total $E_{\rm nonth}$ of 7.9~$\times 10^{30}$ erg for the X2.1 flare using \hsi\ data. In Figure~\ref{f6}(a), we plot the lightcurves of the X1.8 flare with the detector n1 of \fermi/GBM. This detector was the least sunward-facing detector among the four \fermi\ detectors that observed this flare, and it was chosen to reduce the pulse pileup effect. Fitting of X-ray spectra was conducted for 44 intervals throughout the flare (each with an integration time of \sm4~s) and in an energy range of \sm8--300~keV (consisting of about 80 energy bins). The 30--50~keV range was excluded due to a currently inappropriate handling of the iodine K-edge in the response matrix of \fermi's NaI detectors. The thermal plasma and high-energy emissions are modeled with an isothermal (the function $vth$ in SSW) and a nonthermal (the function $thick2$ in SSW) component, respectively. The latter is assumed to be produced by bremsstrahlung emission from electrons in a power-law energy distribution \citep[e.g.,][]{emslie12}. Following \citet{ireland13}, we fix both the break energy and the high energy cutoff to 32~MeV and the spectral index above the break energy to 6 whenever appropriate. Sample fittings at the HXR peaks are shown in Figures~\ref{f6}(b) and (c). Summing the derived nonthermal power over all the fitted intervals, the total $E_{\rm nonth}$ of the X1.8 flare is about 1.6~$\times 10^{31}$ erg. 

We caution that $\Delta E_{\rm free}$ could be up to two times greater than the obtained values, mainly due to the preprocessing applied to the photospheric vector data \citep[e.g.,][]{metcalf08,feng13}. Also, $E_{\rm nonth}$ should be regarded as the lower limit as there exists an uncertainty in determining the low energy cutoff; moreover, $E_{\rm nonth}$ of these two flares is derived using two different instruments, i.e., \hsi\ and \fermi, the systematic difference of which is not yet clear. Therefore, the present results can only suggest that there seems to be no significant difference in the flare energy release between the X2.1 and X1.8 flares. We further note that \fermi/GBM data also show a clear increase in count rates above \sm300~keV in both flares. A detailed analysis of $\gamma$-ray emission is not intended in this paper.

\begin{figure*}
\epsscale{1.17}
\plotone{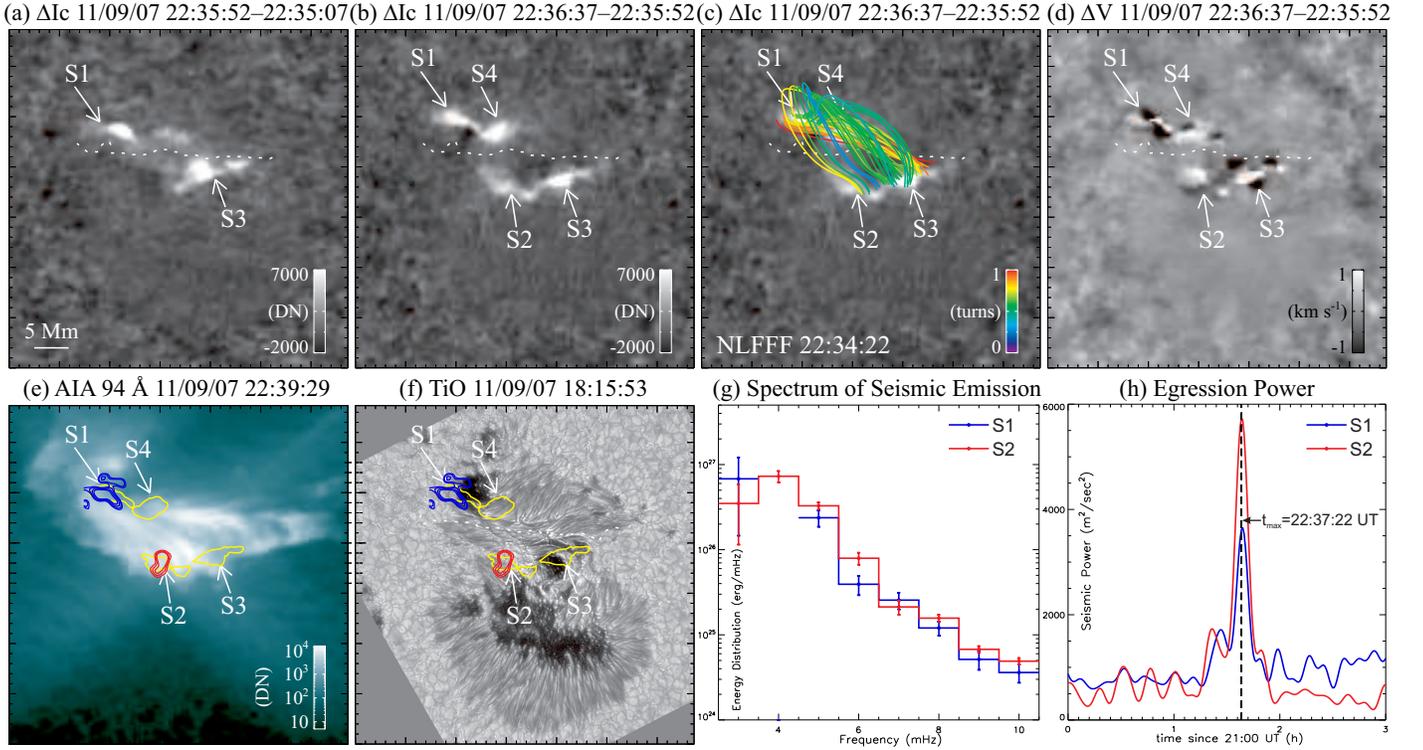}
\caption{Difference images of HMI continuum intensity $\Delta I_c$ (a--c) and the Doppler velocity $\Delta V$ (d), showing four flare impacts S1--S4 in the 2011 September 7 X1.8 flare. Selected representative NLFFF lines are overplotted on (c). The impact sources are also superimposed as yellow contours on an AIA 94~\AA\ image (e) and a BBSO/NST TiO image (f), where the blue and red contours depict the identified two seismic sources (at 6~mHz) at the sites of S1 and S2. The dotted lines are the main segment of the flaring PIL. We show in (g) the seismic spectrum of S1 and S2, and in (h), the time profiles of their egression power at 6~mHz.\vskip 5mm \label{f7}}
\end{figure*}

\begin{figure}
\epsscale{1.17}
\plotone{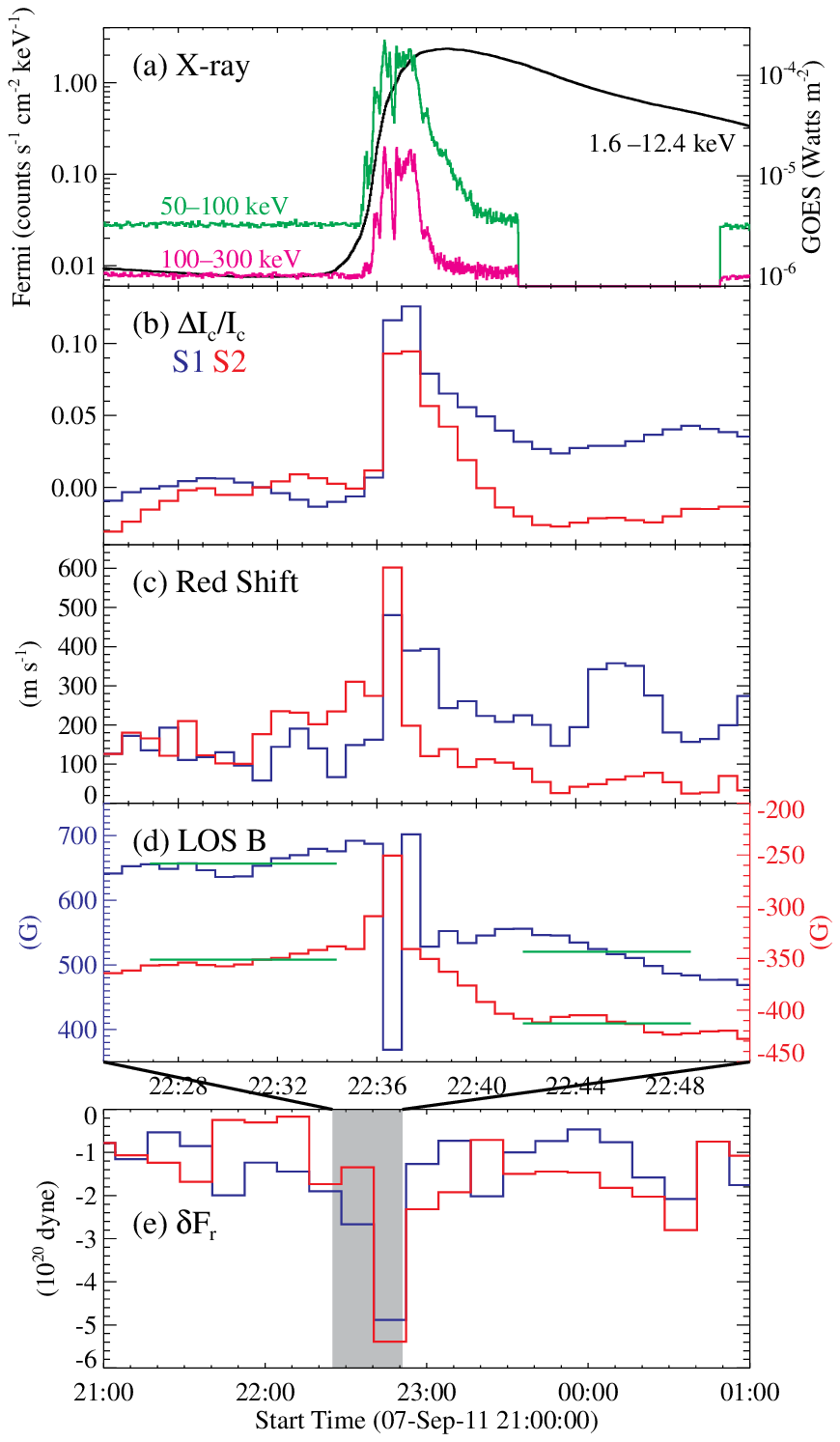}
\caption{Time profiles of X-rays (a) and some properties of the impact sources S1 (blue) and S2 (red) measured at their locations at 22:36:37~UT (Figure~\ref{f7}(b)), including the WL intensity contrast (b), red shift (c), LOS field (d), and the vertical Lorentz-force change derived using the vector field data (e) (a negative value means a downward force change). The green lines in (d) indicate the mean value of LOS magnetic field in the pre- and postflare states. The plotted time period of (a)--(c) is denoted as the gray band in (e). \label{f8}}
\end{figure}

\subsection{Helioseismic Response to Flares} \label{sunquake}
To study the seismicity of our homologous flares, we first map out flare impacts on the photosphere based on the running difference frames of HMI continuum intensity images $\Delta I_c$ and Dopplergrams $\Delta V$ \citep[e.g.,][]{kosovichev11}. We then detect acoustic emission sources by applying the helioseismic holography technique. Within the scope of this paper, our main concerns are the location and relative strength of sunquake sources, and their spatiotemporal relation to flare emissions and magnetic field structure.

We begin with the 2011 September 7 X1.8 flare, which showed a seismic activity \citep{zharkov13}. As a reminder, for this event only HXR time profiles and spectra but no images were available from \fermi/GBM. At the initial HXR spike in 100--300~keV (22:35:52~UT; dotted line in Figure~\ref{f6}(a)), two impact sources S1 and S3 in WL appear near the two ends of the flaring PIL in opposite magnetic polarities (Figure~\ref{f7}(a)). Then around the HXR peak (22:36:37~UT), S1 and S3 become further away from the PIL and two additional impacts S2 and S4 are formed (Figure~\ref{f7}(b)). According to the NLFFF extrapolation (Figure~\ref{f7}(c)), S1 is connected to S2 and S3 to S4, which could also be confirmed by the 94~\AA\ image (Figure~\ref{f7}(e)), with S1 and S3 lying close to the two footpoints of the erupting FR (cf. Figures~\ref{f3}(c) and (d)). These impacts are also characterized by a strong increase of the Doppler velocity up to 1.4~\kms\ (e.g., see Figure~\ref{f7}(d)), indicating a downward flow in the photosphere. Subsequently, the WL emissions show some separation motion away from the PIL as described in the standard flare model, but they rapidly become diffused and weakened after the HXR peak from \sm22:38~UT.

Most intriguingly, despite pronounced WL emission at all the four locations of S1--S4, it is resolved by the helioseismic holography that significant seismic transients are only emitted by S1 and S2 (also see \citealt{zharkov13}), and that the seismic sources are cospatial with S1/S2 at the time around the HXR peak (Figure~\ref{f7}(b)). We clearly demonstrate this spatial correlation and the source location in the photosphere by overplotting the impacts S1--S4 at the HXR peak (yellow) and the identified seismic sources (blue and red) as contours on a corresponding AIA 94~\AA\ image (Figure~\ref{f7}(e)) and an earlier BBSO/NST TiO image (Figure~\ref{f7}(f)). To help understand the exciting mechanism of the sunquake, we also plot the seismic spectra and the time profiles of seismic power of S1/S2 in Figures~\ref{f7}(g) and (h), and compare some additional physical properties of S1 and S2 in Figure~\ref{f8} and Table~\ref{table_1}. Many aspects of these results deserve detailed discussions.

First, several pieces of evidence point to the thick-target heating by electrons as a cause of this sunquake event. These include that (1) the seismic spectra of S1 and S2 have a power-law form and display a similarity over a large frequency band of \sm3--10~mHz (Figure~\ref{f7}(g)). This suggests a common source of accelerated electrons from a coronal reconnection site. These electrons may precipitate into the lower atmosphere along magnetic loops linking S1 and S2, which are indeed both modeled and observed (Figures~\ref{f7}(c) and (e)). (2) The seismic sources S1 and S2 reach their maximum power almost simultaneously at 22:37:22~UT (Figure~\ref{f7}(h)), cotemporal with the peak of HXRs (dashed line in Figure~\ref{f6}(a)). This simultaneity also implies a direct magnetic connection between S1 and S2. (3) The seismic energy of S1 and S2 are estimated to be 4.55~$\times$~10$^{26}$ and 1.53~$\times$~10$^{27}$~erg, respectively, for frequencies above 3~mHz. A stronger seismic emission at S2 could be consistent with the fact that the photospheric vertical field at S2 is much weaker than that at S1 (Table~\ref{table_1}), and hence electron precipitation along S1--S2 may be stronger at S2 due to the magnetic mirroring effect \citep[e.g.,][]{kundu95}.

Second, the WL emissions at S1/S2 peak at 22:37:22~UT (see Figure~\ref{f8}(b)). Thus, they are not only spatially but also temporally correlated with the seismic transients. However, the impact S2 causes a weaker enhancement of WL intensity contrast (9.4\%) than S1 (13\%). Here the image contrast is defined as $\Delta I_c/I_c=(I-I_0)/I_0$, where $I$ is the intensity of the feature of interest, and $I_0$ has a value of $I$ in the preflare state (at 22:31~UT) \citep[e.g.,][]{xu06,kosovichev11}. This casts doubts on the back-warming of the photosphere, which is a secondary effect associated with the WL emission, as the mechanism for the sunquake sources S1 and S2. We conjecture that the initial atmospheric conditions at S1/S2 may allow a direct impingement of electrons to cause seismic emissions.

\begin{deluxetable*}{ccccccccccc}
\tablecolumns{7}
\tablewidth{0pt}
\tablecaption{Physical Properties of Flare Impacts in the 2011 September 7 X1.8 Flare \vskip -2mm \label{table_1}}
\tablehead{\colhead{Photospheric} & \colhead{Sunquake} & \colhead{Acoustic Energy} & \colhead{Peak of} & \colhead{Work by} & \colhead{$|B_r|$} &\colhead{Horizontal Flows\tablenotemark{b}} \\
\colhead{Impacts} & \colhead{Source?} & \colhead{(10$^{27}$ erg)} & \colhead{$\Delta I_c / I_c$ (\%)} & \colhead{$\delta F_r$\tablenotemark{a} (10$^{27}$ erg)} & \colhead{(G)} & \colhead{(\kms)}}
\startdata
S1 & Yes & 0.46 & 13  & 0.36 & 880  & 0.17  \\
S2 & Yes & 1.5  & 9.4 & 0.54 & 330  & 0.061 \\
S3 & No  & --   & 24  & 3.4  & 1300 & 0.15  \\
S4 & No  & --   & 18  & 0.32 & 620  & 0.12 
\enddata
\tablenotetext{a}{Derived using HMI vector magnetograms.}
\tablenotetext{b}{Preflare values.\\}
\end{deluxetable*}

Third, we also estimate the work $W_{LF}$ done by the Lorentz-force change $\delta F_r$ on the interior. Following the derivation of \citet{alvarado12}, $W_{LF} \approx \frac{1}{4} \delta F_r \Delta V \Delta t$, where $\Delta V$ is the red shift transient seen in Dopplergrams and $\Delta t$ is the duration of $\Delta V$. We use two kinds of HMI magnetic field data for the evaluation of $\delta F_r$. (1) The LOS magnetograms have the same 45~s cadence with WL intensity images and Dopplergrams, and show step-like changes of LOS field at impacts S1 and S2 closely associated with the flare (Figure~\ref{f8}(d)). As in \citet{alvarado12}, we ignore the large excursions around the flare impulsive phase that may be affected by flare emissions; instead, we average the LOS field over a 8 minute interval both before (around 22:30:37~UT) and after (around 22:45:14~UT) the flare (denoted by green lines in Figure~\ref{f8}(d)), and take their difference as $\delta B_{\rm los}$. For S1 in the positive field region, $\delta B_{\rm los} \approx -140$~G indicating a decrease of flux, while for S2 in the negative field, $\delta B_{\rm los} \approx -62$~G indicating an increase. Considering the disk location of this AR (N14$^{\circ}$W28$^{\circ}$) and the orientation of the loop S1--S2, the magnetic flux at S1 (S2) region is diskward (limbward). Thus the above LOS field changes are both suggestive of a collapse of the field lines connecting S1--S2 \citep{wang10}, which corresponds to a downward Lorentz-force change. We derive that the \textit{downward} force change is $\delta F_r \approx \frac{1}{8\pi} \int dA |\delta B_{\rm los}^2| \approx 3.4 \times 10^{20}$~dyne at S1 and 1.8$\times 10^{20}$~dyne at S2, respectively. With the result of peak $\Delta V$ around 22:36:37~UT shown in Figure~\ref{f8}(c) and taking $\Delta t \approx 90$~s, we have $W_{LF} \approx 2.5 \times 10^{26}$~erg at S1 and $1.9 \times 10^{26}$~erg at S2. However, these values would be an underestimation because they are only based on the LOS field measurement. (2) We also calculate $\delta F_r$ using vector magnetograms with a 12 minute cadence. The result shown in Figure~\ref{f8}(e) has peaks of $\delta F_r$ around 22:46:22~UT, with which we obtain $W_{LF} \approx 3.6 \times 10^{26}$~erg at S1 and $5.4 \times 10^{26}$~erg at S2. We caution that in deriving $W_{LF}$ above, $\delta F_r$ and $\Delta V$ measured at different times have to be used due to the limitation of observations. Nevertheless we do not expect this approximation to alter the result significantly, assuming that flare-related photospheric magnetic field change tends to be stepwise \citep[e.g.,][]{sudol05}. Since for both impacts S1 and S2, $W_{LF}$ represents a considerable fraction of the seismic energy, the Lorentz-force change may play an important role in producing this sunquake. It also needs to be pointed out that although $\delta F_r$ at S1/S2 amounts only to a few percent of that exerted on the region R2 around the PIL (see Figure~\ref{f1}(f)), the photosphere undergoes a downward perturbation (as indicated by red shifts) predominantly at S1--S4 (Figure~\ref{f7}(d)).

Fourth, it is obvious that S1 is at the poorly developed penumbra of the northern sunspot, and S2 is located just at the edge of a penumbra segment of the largest spot of the AR (Figure~\ref{f7} (f)). In comparison, S3 lies in the umbral region while S4 is in the inner penumbral area. In particular, S3 has a much stronger vertical field than the other three impacts (see Table~\ref{table_1}). As suggested by previous studies \citep[e.g.,][]{donea11,kosovichev11}, wave motions could be restricted by strong magnetic field, which may at least partially explain the absence of seismic emission at S3. In fact, seismic sources are known to predominantly appear in sunspot penumbrae with highly inclined magnetic fields \citep{donea11,lindsey14}. It is however not understood why S4 does not allow the development of a seismic source; perhaps, this is related to the initial conditions of the low atmosphere. We further notice that the region of the impact S2, which induces the strongest seismic emission in this event, has significantly weakest vertical field and photospheric flow right before the flare (see Figure~\ref{f1}(g) and Table~\ref{table_1}). We surmise that a weak surface flow field might also provide a favorable (but not a necessary) condition for the sunquake generation, which needs further observational confirmation and interpretation.

Fifth, it is worthwhile to compare this 2011 September 6 X1.8 flare with the 2011 February 15 X2.2 flare, both of which have multiple sunquake sources and are accompanied by a FR eruption. \citet{zharkov11} found that in the X2.2 flare, two seismic sources are located in the penumbral field at the two footpoints of an erupting FR, and that the acoustic emission precedes the HXR peak by several minutes and is displaced from the strongest HXR source. In the present X1.8 flare, we also see that the seismic sources S1 and S2 are in the penumbral field; however, a noticeable difference is that although the weaker S1 is cospatial with the northern footpoint, the stronger S2 is clearly located away from the southern footpoint of the FR (cf. Figures~\ref{f3}(c)--(d) and \ref{f7}(f)). The afore-described evidence favors the thick-target heating at the HXR peak time caused by the direct precipitation of electrons along the loop S1--S2 as the sunquake driver in the X1.8 flare.

After repeating the analysis for the 2011 September 6 X2.1 flare, we find two bright WL sources (P1 and P2) at the HXR peak time (see the red contours in Figure~\ref{f1}(c)). However, within the detection limit of HMI, no clear seismic sources or expanding wave ripples can be identified. This is intriguing as the X2.1 flare is homologous to the X1.8 flare, showing similar eruption characteristics and energy release. A closer look reveals that P1 and P2 are mainly around the umbral region (cf. Figures~\ref{f0}(a) and \ref{f1}(c)), with a vertical field (as well as flow field) strength close to that of S3. We thus suggest that compared to the X1.8 flare, the lack of a clear seismic signature in the X2.1 flare could be due to the strong vertical field at the location of the flare impacts, which may be unfavorable for the downward energy transfer and hence the generation of seismic waves, when compared to the inclined magnetic fields of penumbral regions \citep{lindsey14}. Additionally, a less well-formed FR and a slower coronal implosion in the X2.1 flare may signify a less amount of momentum deposited into the surface and below.

\begin{deluxetable*}{ccccccccccc}
\tablecolumns{9}
\tablewidth{0pt}
\tablecaption{Physical Properties of 3D Magnetic Field Restructuring \vskip -2mm \label{table_2}}
\tablehead{\colhead{Events} & \colhead{$\Delta E_{\rm free}$\tablenotemark{a}} & \colhead{$E_{\rm nonth}$\tablenotemark{a}} & \multicolumn{2}{c}{Speed (\kms) of Implosions} & \colhead{} & \multicolumn{2}{c}{Distance (Mm) of Implosions} & \colhead{$\Delta B_{\rm h}$} & \colhead{Sunquake}\\
\cline{4-5} \cline{7-8} \\
\colhead{} & \colhead{(10$^{31}$ erg)} & \colhead{(10$^{31}$ erg)} & \colhead{On Surface} & \colhead{Vertical Direction} & & \colhead{On Surface} & \colhead{Vertical Direction} & \colhead{} & \colhead{Association}}
\startdata
2011 Sep 6 X2.1 & 3.5 & 0.79\tablenotemark{b} & 0.26 & 1.5 & & 0.85 & 2.1 & 26\% & No   \\
2011 Sep 7 X1.8 & 4.6 & 1.6                  & 0.17 & 3.0 & & 1.4 & 4.2 & 38\% & Yes   
\enddata
\tablenotetext{a}{Both $\Delta E_{\rm free}$ and $E_{\rm nonth}$ are lower limits.}
\tablenotetext{b}{\citet{feng13}\\}
\end{deluxetable*}

\section{SUMMARY AND DISCUSSION} \label{summary}
We have presented a detailed study of the homologous 2011 September 6 X2.1 and September 7 X1.8 flares concentrating on their magnetic field evolution, energy release, and the helioseismic response. By synthesizing the results of photospheric field change indicated by observations and coronal magnetic field variation suggested by the NLFFF modeling, we have constructed a comprehensive picture of the flare-related 3D magnetic restructuring, which turns out to be consistent with the coronal implosion scenario in the low solar atmosphere. In the aspect of seismicity, we analyzed different flare impacts and the associated seismic sources in the X1.8 flare, which provides clues to the absence of sunquake during the X2.1 flare. The main results are listed in Tables~\ref{table_1} and \ref{table_2} and summarized as follows.

\begin{enumerate}

\item In the photosphere, (1) the flux-weighted centroid separation of the opposite magnetic polarities shrinks dramatically after both flares. We propose that this can be a surface signature of coronal implosions. While the approaching motion has a somewhat higher speed in the X2.1 flare, the total displacement in the X1.8 flare is about a factor of two larger. (2) The horizontal magnetic field rapidly increases around the flaring PIL, but exhibits an asymmetric decrease in the surrounding penumbral regions with a much larger magnitude in the northern area. The WL penumbral structure also exhibits corresponding changes. Both the mean horizontal field strength and inclination angle have larger percentage changes associated with the X1.8 flare.

\item In the corona, (1) the twisted FR lying along the PIL collapses toward the surface after the X2.1 flare, then rises gradually (up to 0.1~\kms) in 1 day reaching a higher altitude, and collapses again after the X1.8 flare. Both the distance and speed of the falling motion in the X1.8 flare are twice as large as those in the X2.1 flare, indicating a more violent implosion and agreeing with a more significant change of the photospheric magnetic field. (2) The FR is attached to the photosphere before the X2.1 flare; in contrast, it is already elevated off the surface before the X1.8 flare. A fuller FR eruption could occur in the latter event, as little signatures of FRs remain afterward. (3) The FR is not symmetric in its central vertical cross section but leans northward at 66\dg\ relative to the surface. Together with the ambient fields, they turn southward rapidly after both flares, echoing the observed asymmetric variations on the surface.

\item The released free magnetic energy and nonthermal electron energy are evaluated using the coronal field extrapolation model and HXR observations, respectively. Due to uncertainties involved in the computation, it can only be concluded that these two homologous flares are energetically similar.

\item Among the four impact sources S1--S4 created by the X1.8 flare on the photosphere, only S1 and S2 spawn seismic emissions. Observational and modeling evidence, including similar seismic spectra, simultaneous excitation, and footpoint asymmetry, all favors the thick-target heating along loops S1--S2 over back-warming as the sunquake mechanism. However, the possibility is not excluded that the downward perturbation by the Lorentz-force change may also contribute. Intriguingly, no sunquake signatures are detected in the homologous X2.1 flare with similar eruption characteristics. A further comparison of magnetic field property among S1--S4 and between the two homologous flares concur with previous studies that seismic emission is more likely to be produced when the flare impact occurs to a penumbral region.

\end{enumerate}

Importantly, our observational and model results portray a coherent picture of implosions in the low corona, in which the central field collapses towards the photosphere while the peripheral field turns to a more vertical configuration, as depicted by \citet{liu05}. Moreover, the implosion process appears to be more abrupt when associated with a fuller FR eruption. Compared to the downward Lorentz-force change resulting from the implosion, the generation of seismic emission sources tends to be more closely related to the precipitation of the flare-accelerated electrons. Nonetheless, our study demonstrates that coronal magnetic field structure and the photospheric magnetic field property of the flare impact locations are helpful in understanding the triggering mechanism of sunquakes. Extended studies on the spatiotemporal evolution involving 3D magnetic field restructuring, particle acceleration, and helioseismic response hold promise of shedding further light on their causal relationship.

\acknowledgments
We thank the \sdo\ team for the magnetic and EUV/UV data. \sdo\ is the first satellite under the Living with a Star (LWS) program at NASA. We acknowledge the use of the \fermi\ Solar Flare Observations facility funded by the \fermi\ GI program. We also thank the BBSO/NST team for the TiO image and the anonymous referee for helpful comments. C.L. is grateful to Dr. Yan Xu for helpful discussions on the thick-target heating. C.L., N.D., and H.W. were supported by NASA under grants NNX13AF76G and NNX13AG13G issued through the LWS program and NNX14AC12G, and by NSF under grant AGS 1408703. J.L. was supported by the Brainpool Program 2014 of KOFST. T.W. was supported by DLR grant 50 OC 0904 and DFG grant WI 3211/2-1. C.W.J was supported by the National Natural Science Foundation of China 41204126. Y.S. acknowledges the Austrian Science Fund (FWF) P24092-N16 and FWF P27292-N20.

\end{document}